# Electronic structure and spin dynamics of $A$Co$_2$As$_2$ ($A$=Ba, Sr, Ca)

Huican Mao and Zhiping Yin[*]

*Department of Physics and Center for Advanced Quantum Studies, Beijing Normal University, Beijing 100875, China*

**Abstract**

   The electronic structures, charge and spin dynamics of the cobalt pnictide compounds $A$Co$_2$As$_2$ ($A$=Ba, Sr, Ca) in the paramagnetic state are investigated by using density functional theory combined with dynamical mean-field theory. In contrast to their iron counterparts, these cobalt pnictide compounds have three-dimensional electronic structures and strong ferromagnetic low-energy spin excitations. The Co 3$d$ $e_g$ orbitals dominate the electronic states around the Fermi level and have stronger electronic correlation strength than the Co 3$d$ $t_{2g}$ orbitals. The overall electronic correlation strength is much weaker than that in the iron arsenides; however, the most strongly correlated Co 3$d$ $x^2$-$y^2$ orbital, especially in CaCo$_2$As$_2$, has electronic correlation strength comparable to Fe 3$d$ $t_{2g}$ orbitals in iron arsenides. $A$Co$_2$As$_2$ ($A$=Ba, Sr, Ca) shows similar electronic structures where a conduction band of primarily Co 3$d$ $x^2$-$y^2$ orbital character is close to a Van Hove singularity around the Brillouin-zone corner, which promotes ferromagnetic low-energy spin excitations. Originated from its increased nearest-neighbor Co-Co distance and significantly reduced As height from the Co plane, the strong electronic correlation strength and close proximity to the Van Hove singularity of the Co 3$d$ $x^2$-$y^2$ orbital in CaCo$_2$As$_2$ is responsible for its unique A-type antiferromagnetic order observed in experiments. In comparison, despite substantial ferromagnetic low-energy spin excitations, BaCo$_2$As$_2$ and SrCo$_2$As$_2$ remain paramagnetic down to very low temperature because the Co 3$d$ $x^2$-$y^2$ orbital has weaker electronic correlation strength and is further away from the Van Hove singularity.

I.  **Introduction**

The discovery of superconductivity at 26 K in La(O$_{1-x}$F$_x$)FeAs (x = 0.05 - 0.12) [1] creates tremendous activity in the scientific community. So far, $A$Fe$_2$As$_2$ ($A$=Ba, Sr, Ca) 122 systems have been extensively studied because of the availability of large single crystals [2-6]. They all exhibit simultaneous structural and stripe antiferromagnetic (AFM) transition at the Neel temperature of 143 K (BaFe$_2$As$_2$) [7], 203 K (SrFe$_2$As$_2$) [8] and 167 K (CaFe$_2$As$_2$) [9]. The cobalt arsenides $A$Co$_2$As$_2$ ($A$= Ba, Sr, Ca) with a full substitution of Fe by Co have received increased attention recently due to their close relationship to the parent compounds of the $A$Fe$_2$As$_2$ superconductor family.

The $A$Co$_2$As$_2$ compounds appear to exhibit very different behavior as compared to $A$Fe$_2$As$_2$ ($A$=Ba, Sr, Ca). Experimental results show no structure phase transition and superconductivity down to 2 K for $A$Co$_2$As$_2$ ($A$=Ba, Sr, Ca) [10-12]. Neutron-diffraction measurements on SrCo$_2$As$_2$ show no evidence for long-rang magnetic ordering above 2 K [11]. The isoelectronic BaCo$_2$As$_2$ exhibits the same feature, as suggested by magnetic susceptibility, electrical resistivity and specific-heat measurements [13]. In contrast to BaCo$_2$As$_2$ and SrCo$_2$As$_2$, CaCo$_2$As$_2$ undergoes an A-type AFM order below a sample-dependent Neel temperature of 52 -76 K [14], where the magnetic moments of Fe atoms order ferromagnetically in the *ab*-plane and antiferromagnetically along the *c*-axis. Both the A-type AFM structure and the positive Weiss temperature in the Curie-Weiss law indicate that the dominant interactions in CaCo$_2$As$_2$ are ferromagnetic (FM) [10,14]. In SrCo$_2$As$_2$, nuclear magnetic resonance (NMR) measurements and inelastic neutron scattering suggest that both FM and stripe-type AFM fluctuations coexist, which is also supported by density functional theory (DFT) calculations, where the $q$-dependent static susceptibilities $\chi(q)$ at both the FM and stipe AFM in-plane wave vectors are enhanced [15,16]. Sefat *et al.* reported that the experimental measurements of magnetic, resistivity and thermal properties, combined with DFT calculations, show that BaCo$_2$As$_2$ is a highly renormalized paramagnet [13]. Moreover, NMR techniques indicate that both the Knight shift $^{75}$K and $1/T_1T$ increase toward $T$ = 0 because of FM spin correlations in BaCo$_2$As$_2$.

Furthermore, van Roekeghem *et al.* combined the screened exchange and dynamical mean-field-theory (SE + DDMFT) scheme to calculate the spectral function of $BaCo_2As_2$ and showed that the ferromagnetic instability is absent in this compound [17]. Nevertheless, extensive data analysis does not entirely rule out the possibility of AFM spin correlations [18]. Additionally, angle-resolved photoemission spectroscopy data and electronic structure calculations on $BaCo_2As_2$ [19,20] and $SrCo_2As_2$ [11] reveal a complex multiband Fermi surface without clear nesting features.

However, despite extensive research, it is still not clear why only $CaCo_2As_2$ has an A-type AFM transition while $BaCo_2As_2$ and $SrCo_2As_2$ exhibit paramagnetism down to 2 K. Furthermore, there is no consistent conclusion on which kind of magnetic fluctuation (AFM or FM) is dominant in $ACo_2As_2$ (*A*=Ba, Sr, Ca).

## II. Computational details

In this paper, we use fully charge self-consistent density functional theory plus dynamical mean-field theory (DFT+DMFT) [21,22] to theoretically study the archetypical cobalt pnictide compounds $ACo_2As_2$ (*A*=Ba, Sr, Ca) in the paramagnetic (PM) state. The DFT part is based on the linearized augmented plane-wave method as implemented in WIEN2K [23]. We use the Perdew–Burke–Ernzerhof exchange correlation functional. Hubbard $U = 5.0$ eV and Hund's coupling $J = 0.8$ eV are used in the calculations, consistent with previous calculations [24-26]. The formula $U(n - 1/2) - J(n - 1)/2$ (*n* is the nominal occupation of Co 3*d* electrons) is used to subtract the double counting. The impurity problem in DFT+DMFT calculations is solved using continuous time quantum Monte Carlo (CTQMC) method [27,28] at temperature $T = 72.5$ K. The dynamical spin structure factor is calculated using the method described in detail in Ref. [25]. We use the experimentally determined crystal structures including the internal atomic positions [10-12].

## III. Results

After achieving full charge self-consistency, we compute the momentum resolved spectra function (band structure), density of states (DOS), Fermi surface (FS), optical conductivity and spin excitation spectra. We find that the overall mass enhancement of the Co 3$d$ electrons is much smaller in $A$Co$_2$As$_2$ ($A$=Ba, Sr, Ca) than that of Fe 3$d$ electrons in BaFe$_2$As$_2$ and the Co 3$d$ $e_g$ orbitals have larger mass enhancement than the $t_{2g}$ orbitals, in contrast to the Fe 3$d$ orbitals in Fe-based compounds where the $t_{2g}$ orbitals have larger mass enhancement than the $e_g$ orbitals. The Co 3$d$ $x^2$-$y^2$ orbital is the most strongly correlated orbital whereas the 3$d$ $xy$ orbital is the least correlated orbital. The calculated dynamical spin structure factor shows that CaCo$_2$As$_2$ has strong ferromagnetic spin excitations which diverge at low energy and temperature and are responsible for the observed A-type AFM order in CaCo$_2$As$_2$. In contrast, although BaCo$_2$As$_2$ and SrCo$_2$As$_2$ also have dominate ferromagnetic spin excitations, the strength is much weaker than that in CaCo$_2$As$_2$. Therefore, they do not undergo a magnetic transition down to very low temperature and remain paramagnetic, consistent with experiments. We further trace the difference in the spin excitations among $A$Co$_2$As$_2$ ($A$=Ba, Sr, Ca) to differences in the electronic correlation strength of the Co 3$d$ $x^2$-$y^2$ orbitals and the electronic structures around the Fermi level which originate primarily from the different As heights in these cobalt pnictides compounds.

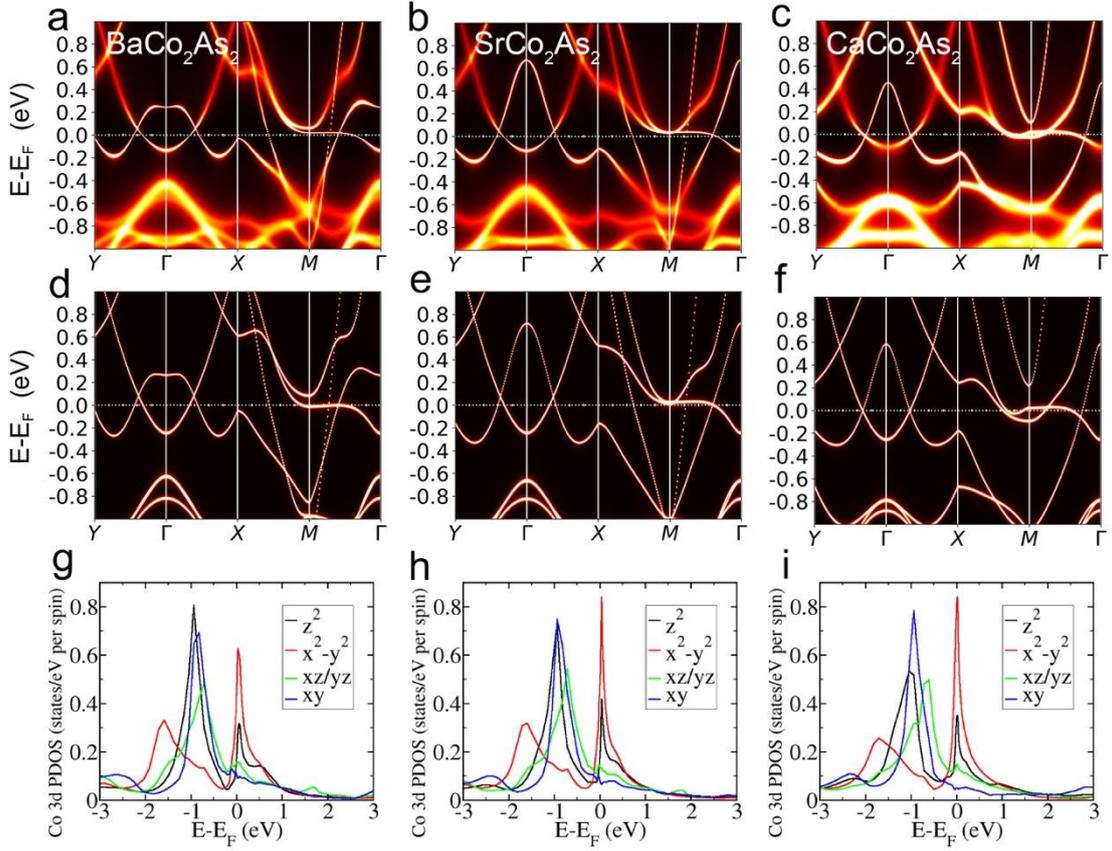

Fig.1 The electronic band structure (a-f) and Co 3$d$ partial density of states (g-i) of BaCo$_2$As$_2$ (a, d, g), SrCo$_2$As$_2$ (b, e, h), and CaCo$_2$As$_2$ (c, f, i) obtained by DFT+DMFT calculations (a-c), (g-i) and standard DFT calculations (d-f) in the paramagnetic state.

### Band structure and density of states

Figure 1 shows the electronic band structure and Co 3$d$ partial density of states of $A$Co$_2$As$_2$ ($A$=Ba, Sr, Ca) based on DFT+DMFT (Fig.1a-c, g-i) and DFT calculations (Fig.1d-f), respectively. The conduction bands are mostly of Co 3$d$ $e_g$ orbital character as evident in the Co 3$d$ partial DOS shown in Fig. 1g-i, in contrast to the iron arsenide compounds. The familiar 3$d$ $t_{2g}$ bands giving rise to the hole (electron) Fermi-surface pockets around the Brillouin-zone center (corner) Γ (M) point in the iron arsenides locate 0.6-0.8 eV below the Fermi level, indicating a somewhat rigid-band shift of the Fermi level due to substitution of Fe by Co.

The overall electronic structures of $A$Co$_2$As$_2$ ($A$=Ba, Sr, Ca) are similar. The main difference lies in a flat conduction band of interest near the Fermi level around the M point. This band gives rise to a sharp peak in the partial DOS of the Co 3$d$ $x^2$-$y^2$ orbital [Fig.1g-i] signaling this band of Co 3$d$ $x^2$-$y^2$ orbital character is close to a Van Hove singularity. The proximity to a Van Hove singularity near the Fermi level leads to high density of states at the Fermi level and promotes ferromagnetic low-energy spin excitations according to the Stoner criterion. The peak position in the partial DOS of the Co 3$d$ $x^2$-$y^2$ orbital is very close to the Fermi level in CaCo$_2$As$_2$ whereas it is about 35 meV above the Fermi level in BaCo$_2$As$_2$ and SrCo$_2$As$_2$, suggesting CaCo$_2$As$_2$ is in close proximity to the Van Hove singularity while BaCo$_2$As$_2$ and SrCo$_2$As$_2$ are further away from the Van Hove singularity, which is a main reason that CaCo$_2$As$_2$ has much stronger ferromagnetic low-energy spin excitations than BaCo$_2$As$_2$ and SrCo$_2$As$_2$ as shown later in the paper. Moreover, in CaCo$_2$As$_2$, two conduction bands around the M point cross the Fermi level and each other to form two separated dumbbell-shape hole-electron pockets, further promoting ferromagnetic (zero-momentum-transfer) low-energy spin excitations.

The absence of ferromagnetism in BaCo$_2$As$_2$ has been discussed previously in the context of SE + DDMFT calculations [17], where it is argued that the long-range exchange interaction reduces the density of states at the Fermi level and avoids the ferromagnetic order according to the Stoner criterion. Here we show that without taking into account the long-range exchange interaction, correlation effects due to local Coulomb interaction already avoid the ferromagnetic ordering in BaCo$_2$As$_2$.

The DFT band structures are also similar to the corresponding DFT+DMFT band structures, subject to renormalization of the bandwidth and some shift of the band positions. The bandwidths of the Co 3$d$ bands around the Fermi level do not vary much among the cobalt pnictide compounds. For example, the DFT+DMFT (DFT) parabolic band around Γ point (along the Y-Γ-X path) giving rise to an electron Fermi surface pocket is mainly of Co 3$d$ $x^2$-$y^2$ orbital character and has a bandwidth of 1.25 (1.55), 1.33 (1.66), and 1.32 (1.71) eV in BaCo$_2$As$_2$, SrCo$_2$As$_2$, and CaCo$_2$As$_2$, respectively. This results in overall band renormalization factors of 1.23, 1.25 and 1.29 for the Co 3$d$ $x^2$-$y^2$ orbital in BaCo$_2$As$_2$, SrCo$_2$As$_2$, and CaCo$_2$As$_2$, respectively, which are significantly smaller than the corresponding mass enhancement of above 2 at the Fermi level calculated by $m^*/m_{band} = 1 - \partial\Sigma'(\omega=0)/\partial\omega$ from the quasiparticle self-energy $\Sigma'(\omega)$ as shown in Table I, suggesting significant nonuniform renormalization of the band at different energies.

The overall mass enhancement of the Co 3$d$ orbitals in $A$Co$_2$As$_2$ ($A$=Ba, Sr, Ca) are below 2 as shown in Table I, indicating much weaker electronic correlation effects than iron arsenide compounds, for which mass enhancement of about 3 was reported [24,25]. This is mainly due to an average increase of 0.2 of the Co 3$d$ orbital occupation because of the substitution of Fe with Co. The Co 3$d$ $e_g$ orbitals have stronger electronic correlation strength than the Co 3$d$ $t_{2g}$ orbitals, in strong contrast to the iron arsenide compounds where the Fe 3$d$ $t_{2g}$ orbitals have stronger electronic correlation strength than the Fe 3$d$ $e_g$ orbitals. The Co 3$d$ $x^2$-$y^2$ orbital has the strongest electronic correlation strength among all the Co 3$d$ orbitals, and in CaCo$_2$As$_2$ its electronic correlation strength is comparable to that of the Fe 3$d$ $t_{2g}$ orbitals in iron arsenides [25]. The stronger electronic correlation strength and higher density of states due to closer proximity to the Van Hove singularity of the Co 3$d$ $x^2$-$y^2$ orbital in CaCo$_2$As$_2$ than in BaCo$_2$As$_2$ and SrCo$_2$As$_2$ result in much stronger ferromagnetic low-energy spin excitations in CaCo$_2$As$_2$ than in BaCo$_2$As$_2$ and SrCo$_2$As$_2$ as shown below.

Table I. Mass enhancement obtained from the quasiparticle self-energy and the CTQMC averaged orbital occupation of Co 3$d$ orbitals from the charge self-consistent DFT+DMFT calculations.

|  | Mass enhancement | | | | Orbital occupation | | | |
| --- | --- | --- | --- | --- | --- | --- | --- | --- |
|  | $z^2$ | $x^2$-$y^2$ | xz/yz | xy | $z^2$ | $x^2$-$y^2$ | xz/yz | xy |
| BaCo$_2$As$_2$ | 1.73 | 2.05 | 1.61 | 1.54 | 1.45 | 1.31 | 1.49 | 1.55 |
| SrCo$_2$As$_2$ | 1.76 | 2.16 | 1.59 | 1.52 | 1.43 | 1.29 | 1.50 | 1.56 |
| CaCo$_2$As$_2$ | 1.64 | 2.53 | 1.46 | 1.33 | 1.45 | 1.30 | 1.50 | 1.57 |

*Fermi surfaces*

The DFT+DMFT calculated three-dimensional FSs for $A$Co$_2$As$_2$ ($A$=Ba, Sr, Ca) are plotted in Fig. 2 in the PM tetragonal Brillouin zone. The corresponding two dimensional FS cuts in the $k_z$ = 0 and 2$\pi$/c plane are displayed in the unfolded Brillouin zone of the one-Co unit cell. As shown in Fig.2(a-c), the FS has multi-sheets which are large and quite three dimensional, in contrast to their Fe counterparts and consistent with the anisotropy of the measured resistivity [13] and the calculated optical conductivity shown below. As expected from the above analysis of the band structure in the vicinity of $E_F$, the FS topology of $A$Co$_2$As$_2$ ($A$=Ba, Sr, Ca) is quite different from that of the Fe pnictides. For example, the more or less circular hole FS pockets around the Γ point in the $k_z = 0$ plane for BaFe$_2$As$_2$ [25] are replaced by star-shape electron FS pockets (rather than hole pockets) in $A$Co$_2$As$_2$ ($A$=Ba, Sr, Ca), thus removing the electron-hole quasinesting scattering from $\boldsymbol{k} = (0, 0)$ to $(1, 0)$ in cobalt pnictide materials, as shown in Fig.2d-i. In addition, we find that CaCo$_2$As$_2$ shows two separated dumbbell-shape FS pockets at $\boldsymbol{k} = (1, 0)$ in the $k_z = 0$ plane, in contrast to an integrated dumbbell-shape FS pocket in BaCo$_2$As$_2$ and SrCo$_2$As$_2$. Similarly, we find that the FS pockets around $\boldsymbol{k} = (0, 0)$ in the $k_z = 2\pi/c$ plane are surrounded by two sets of four smaller FS pockets and exhibit changes of size and shape among $A$Co$_2$As$_2$ ($A$=Ba, Sr, Ca). At $\boldsymbol{k} = (1, 1)$, the shape of the FS pocket in the $k_z = 2\pi/c$ plane alters from star shape to square shape to circle shape in these cobalt pnictide compounds. In particular, comparing with BaCo$_2$As$_2$ and SrCo$_2$As$_2$, CaCo$_2$As$_2$ manifests a distinctive feature, where it has extra FS pockets at $\boldsymbol{k} = (1, 1)$ in the $k_z = 0$ plane and at $\boldsymbol{k} = (1, 0)$ in the $k_z = 2\pi/c$ plane, respectively.

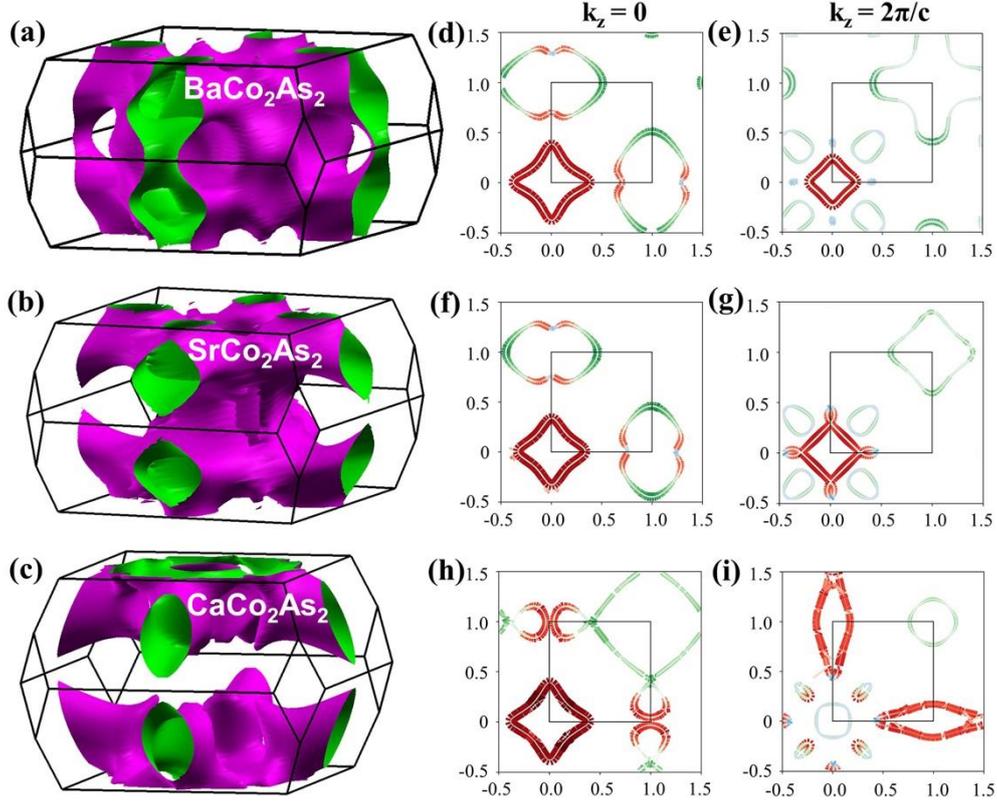

Fig.2 Three-dimensional Fermi surfaces for (a) BaCo$_2$As$_2$, (b) SrCo$_2$As$_2$ and (c) CaCo$_2$As$_2$ plotted in the PM tetragonal Brillouin zone obtained by DFT+DMFT calculations. The corresponding two-dimensional FS cuts in $k_z = 0$ and $2\pi/c$ plane of BaCo$_2$As$_2$ (d, e), SrCo$_2$As$_2$ (f, g), and CaCo$_2$As$_2$ (i, j) in the first Brillouin zone of the one-Co unit cell. The black, red, green, and blue colors denote the Co $3d\ z^2$, $x^2$-$y^2$, $xz/yz$, and $xy$ orbital characters, respectively. In panels (d-i), the units of the reciprocal-lattice vector $k_x$ and $k_y$ are $\sqrt{2}\pi/a$.

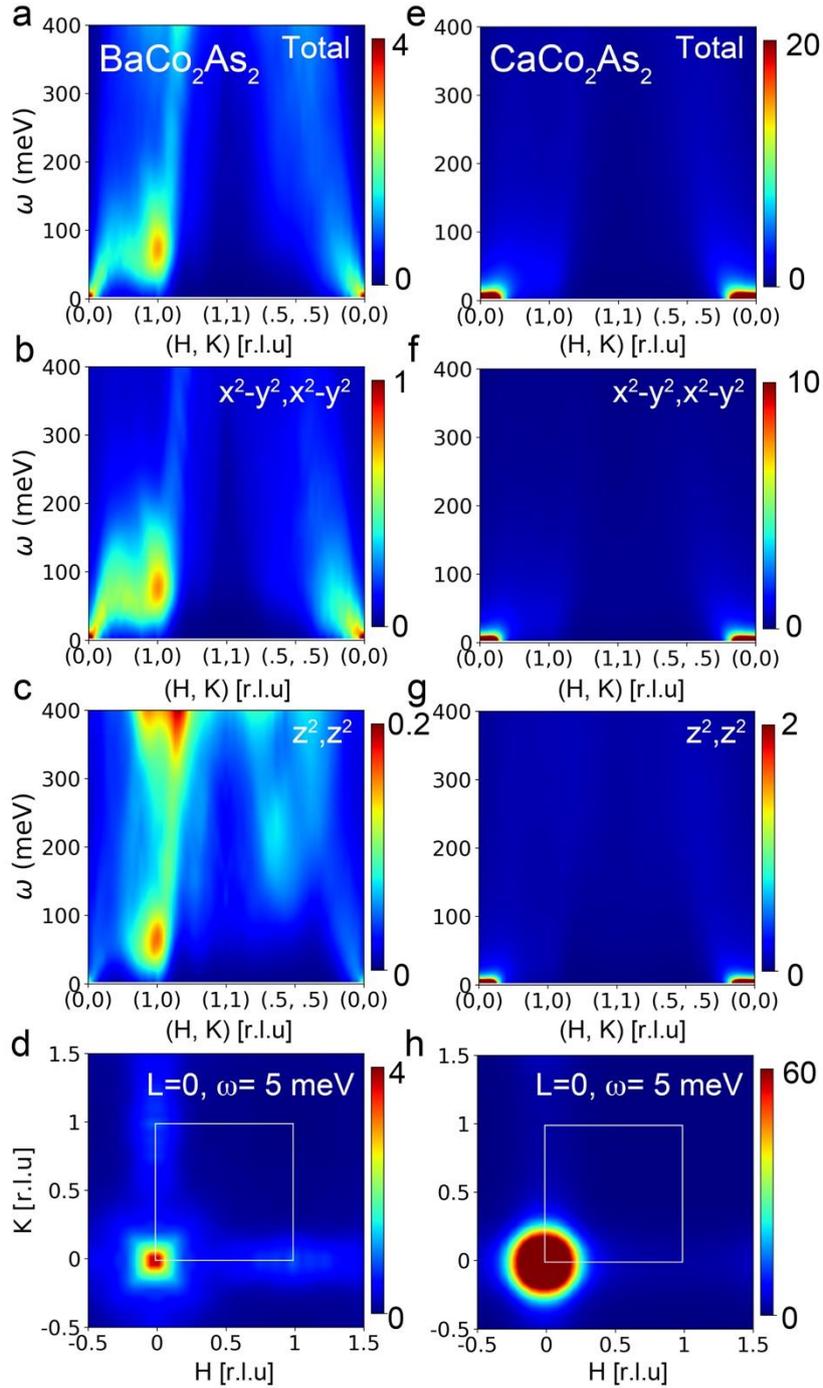

Fig.3 Dynamic spin structure factor $S(q,\omega)$ for BaCo$_2$As$_2$ (left column) and CaCo$_2$As$_2$ (right column). From top to bottom are the total $S(q,\omega)$ (a, e) and the diagonal component $S_{\alpha,\alpha}(q,\omega)$ ($\alpha$ is the orbital index) of Co 3d $x^2$-$y^2$ orbital (b, f) and $z^2$ orbital (c, g) along the high-symmetry path (0, 0)→(1,0)→(1,1)→(0,0), and (d, h) the 2D plot in the L=0 plane with constant energy $\omega = 5\ meV$. The $S(q,\omega)$ of SrCo$_2$As$_2$ is very similar to that of BaCo$_2$As$_2$ thus is not shown. The wave vector is defined in terms of the reciprocal lattice of the one-Co unit cell.

*Spin dynamics*

The calculated dynamical spin structure factor $S(q,\omega) = \chi''(q,\omega)/(1-exp(-\hbar\omega/k_BT))$, which can be measured by inelastic neutron-scattering experiments [29], is shown in Fig.3 for BaCo$_2$As$_2$ (left column) and CaCo$_2$As$_2$ (right column). The dynamical spin structure factor of SrCo$_2$As$_2$ is similar to that of BaCo$_2$As$_2$ and hence is not shown. As is clear from Fig.3a & e, the low-energy spin excitations are much weaker in BaCo$_2$As$_2$ (and SrCo$_2$As$_2$) than in CaCo$_2$As$_2$. Two broad maxima appear at *q* = (0, 0) and the spin-density wave ordering vector *q* = (1, 0) in BaCo$_2$As$_2$ (see Fig. 3a-d), corresponding to FM spin excitations and stripe AFM spin excitations, respectively. In contrast, the one-order-of-magnitude stronger low-energy spin excitation in CaCo$_2$As$_2$ is mainly concentrated on the FM ordering vector *q* = (0, 0) (see Fig. 3e-h), which is expected to diverge at zero energy and low temperature, leading to ferromagnetic ordering of the Co magnetic moments in the Co plane.

We further decompose the dynamical spin structure factor $S(q,\omega) = \sum_{\alpha,\beta} S_{\alpha,\beta}(q,\omega)$ into different orbital contributions, where $\alpha$ and $\beta$ are orbital indices. We find that the low-energy spin excitations are mainly contributed by the Co 3*d* $e_g$ orbitals, in strong contrast to iron arsenides where the low-energy spin excitations are mainly contributed by Fe 3*d* $t_{2g}$ orbitals [30,31]. As shown in Fig. 3b-c and Fig. 3f-g, the diagonal component $S_{\alpha,\alpha}(q,\omega)$ of the Co 3*d* $x^2$-$y^2$ orbital is five times that of the 3*d* $z^2$ orbital, whereas the intensities of the Co 3*d* $t_{2g}$ orbitals are negligibly small. Therefore, the low-energy spin excitations are primarily contributed by the Co 3*d* $x^2$-$y^2$ orbital due to the aforementioned reasons. Furthermore, the intensity of the Co 3*d* $x^2$-$y^2$ orbital in CaCo$_2$As$_2$ is one order of magnitude larger than that in BaCo$_2$As$_2$ (Fig.3b, f) due to the stronger electronic correlation strength and closer proximity to the Van Hove singularity of the Co 3*d* $x^2$-$y^2$ orbital in CaCo$_2$As$_2$ than in BaCo$_2$As$_2$ as discussed above. This explains why CaCo$_2$As$_2$ exhibits unique A-type AFM order at low temperature while BaCo$_2$As$_2$ and SrCo$_2$As$_2$ remain paramagnetic down to very low temperature (~ 2 K).

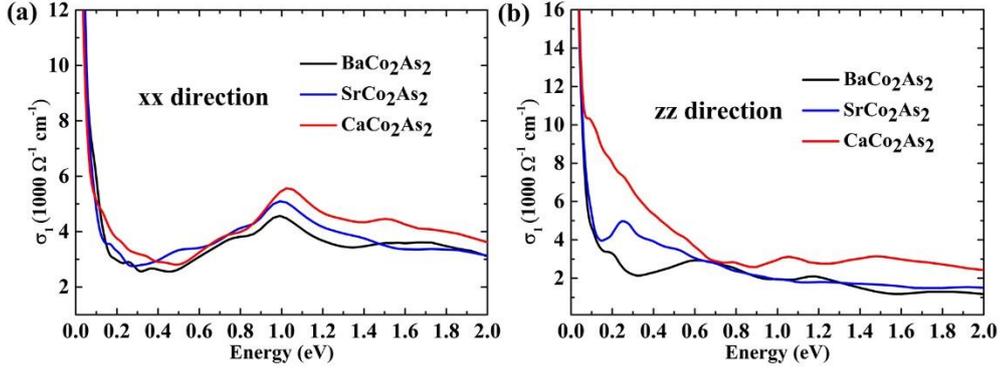

Fig.4 The (a) in-plane and (b) out-of-plane optical conductivity of $A$Co$_2$As$_2$ ($A$=Ba, Sr, Ca) in the PM state obtained by DFT+DMFT calculations.

*Optical conductivity*

Fig.4 shows the in-plane and out-of-plane optical conductivity of $A$Co$_2$As$_2$ ($A$=Ba, Sr, Ca) in the PM state calculated by DFT+DMFT. In Fig.4a, a broad peak, due to interband transitions, is centered around 1.0 eV for in-plane optical conductivity in $A$Co$_2$As$_2$ ($A$=Ba, Sr, Ca) compounds. In contrast, the out-of-plane optical conductivity does not show a clear characteristic peak, whereas it does exhibit obvious differences among $A$Co$_2$As$_2$ ($A$=Ba, Sr, Ca) compounds. Near zero frequency, the optical conductivity spectrum in the PM phase shows Drude-like behavior as shown in Fig.4, where the out-of-plane Drude weight is substantially larger for the cobalt pnictide compounds. The plasma frequencies $\omega_{p,xx}$ and $\omega_{p,zz}$ along the $x$ and $z$ directions for $A$Co$_2$As$_2$ ($A$=Ba, Sr, Ca) and BaFe$_2$As$_2$, obtained from both DFT+DMFT and standard DFT calculations, are shown in Table II. The plasma frequencies from DFT+DMFT calculations are obtained by fitting the calculated low-energy optical conductivity to the Drude function $\sigma_1(\omega) = \frac{\omega_p^2 \Gamma}{\omega^2 + \Gamma^2}$ where $\omega_p$ is the plasma frequency and $\Gamma$ is the scattering rate. The in-plane plasma frequency $\omega_{p,xx}$ is obviously larger in $A$Co$_2$As$_2$ ($A$=Ba, Sr, Ca) than in BaFe$_2$As$_2$, especially for DFT+DMFT calculations. The average mass enhancement $\frac{m^*}{m_{band}} = \omega_p(\text{DFT})^2 / \omega_p(\text{DMFT})^2$ is much smaller for the cobalt pnictide compounds, indicating weaker electronic correlations, consistent with the estimation from the electronic bandwidth, the quasiparticle self-energy, and experimental results.

The in-plane and out-of-plane conductivity anisotropy, defined as $\sigma_{1,x}/\sigma_{1,z} = \omega_{p,xx}^2/\omega_{p,zz}^2$ (assuming the same scattering rate at low temperature), is above 10 in BaFe$_2$As$_2$ as shown in Table II, suggesting strong in-plane and out-of-plane anisotropies and semi-two-dimensional electronic structures in BaFe$_2$As$_2$. In contrast, the in-plane and out-of-plane conductivity anisotropies are around unity in $A$Co$_2$As$_2$ ($A$=Ba, Sr, Ca). Therefore, $A$Co$_2$As$_2$ ($A$=Ba, Sr, Ca) has three-dimensional electronic structures as shown above. In particular, CaCo2As2 has better conductivity along the $c$ axis than in the $ab$ plane.

Table II. The in-plane and out-of-plane plasma frequency ($\omega_{p,xx}$, $\omega_{p,zz}$, respectively), the in-plane and out-of-plane conductivity anisotropy $\sigma_{1,x}/\sigma_{1,z} = \omega_{p,xx}^2/\omega_{p,zz}^2$ and the average mass enhancement along $xx$ and $zz$ directions in $A$Co$_2$As$_2$ ($A$=Ba, Sr, Ca) and BaFe$_2$As$_2$ from Ref. [24, 25].

| | DFT $\omega_{p,xx}$ (eV) | DFT $\omega_{p,zz}$ (eV) | DMFT $\omega_{p,xx}$ (eV) | DMFT $\omega_{p,zz}$ (eV) | DFT $\sigma_{1,x}/\sigma_{1,z}$ | DMFT $\sigma_{1,x}/\sigma_{1,z}$ | Mass enhancement $xx$ | Mass enhancement $zz$ |
|---|---|---|---|---|---|---|---|---|
| BaCo$_2$As$_2$ | 3.66 | 3.37 | 2.77 | 3.06 | 1.18 | 0.82 | 1.75 | 1.21 |
| SrCo$_2$As$_2$ | 3.54 | 3.45 | 2.95 | 2.96 | 1.05 | 0.99 | 1.44 | 1.36 |
| CaCo$_2$As$_2$ | 2.87 | 3.57 | 2.37 | 3.20 | 0.65 | 0.55 | 1.47 | 1.24 |
| BaFe$_2$As$_2$ | 2.63 | 0.77 | 1.60 | 0.5 | 11.7 | 10.2 | 2.7 | 2.4 |

## IV. Discussions

The dimensionality of the electronic structures and the electronic correlation strength are closely connected to the crystal structure parameters [25]. In Table III, we collect the $c$-lattice constant which is twice the interlayer distance of the neighboring Co (Fe) layers, the bond length of Co-Co (Fe-Fe) and Co-As (Fe-As), and the anion As height from the Co (Fe) layer in $A$Co$_2$As$_2$ ($A$=Ba, Sr, Ca) [10-12] and BaFe$_2$As$_2$ [32]. The substantial reduction of the interlayer distance in the cobalt arsenides from BaFe$_2$As$_2$,

~20% in CaCo$_2$As$_2$, is responsible for the significantly increased three-dimensional electronic structures and decreased in-plane and out-of-plane anisotropies in the cobalt arsenides shown above. While the nearest-neighbor Co-Co and Fe-Fe distances do not differ much, the As height in the cobalt arsenides is substantially lower than that in BaFe$_2$As$_2$, leading to a strongly distorted As$_4$ tetrahedron away from its perfect shape and different crystal field splitting, which reduces the electronic correlation strength of the 3d $t_{2g}$ orbitals and promotes the electronic correlation strength of the 3d $e_g$ orbitals [25]. Among $A$Co$_2$As$_2$ ($A$=Ba, Sr, Ca), CaCo$_2$As$_2$ has the largest nearest-neighbor Co-Co bond length and lowest As height, therefore its Co 3d $x^2$-$y^2$ orbital has the strongest electronic correlation strength and plays a crucial role in its unique A-type AFM order at low temperature.

Table III. The $c$-lattice constant, bond length of Co-Co (Fe-Fe) and Co-As (Fe-As), and the anion As height from Co (Fe) layer in $A$Co$_2$As$_2$ ($A$=Ba, Sr, Ca) [10-12] and BaFe$_2$As$_2$ [32].

|  | $c$ (Å) | Co-Co (Å) | Co-As (Å) | $h_{\text{Co-As}}$ (Å) |
|---|---|---|---|---|
| BaCo$_2$As$_2$ | 12.670 | 2.798 | 2.356 | 1.278 |
| SrCo$_2$As$_2$ | 11.773 | 2.791 | 2.352 | 1.280 |
| CaCo$_2$As$_2$ | 10.273 | 2.817 | 2.327 | 1.204 |
| BaFe$_2$As$_2$ | 13.017 | 2.802 | 2.403 | 1.360 |

## V.  Conclusions

While the cobalt arsenides $A$Co$_2$As$_2$ ($A$=Ba, Sr, Ca) are isostructural to their iron counterparts $A$Fe$_2$As$_2$, they show quite different (three-dimensional) electronic structures and (strong ferromagnetic) spin excitations. Except the 3d $x^2$-$y^2$ orbital, the mass enhancement of the Co 3d orbitals is below 2, indicating much weaker electronic correlation strength than that in the corresponding iron compounds. The Co 3d $e_g$ orbitals dominate the electronic states around the Fermi level and have stronger electronic correlation strength than the Co 3d $t_{2g}$ orbitals, in contrast to the iron arsenide compounds where the Fe 3d $t_{2g}$ orbitals have stronger electronic correlation strength than the Fe 3d $e_g$ orbitals [25]. The Co 3d $x^2$-$y^2$ orbital has the strongest electronic correlation strength among all the Co $3d$ orbitals, especially in $CaCo_2As_2$, which is comparable to the correlation strength of the Fe $3d$ $t_{2g}$ orbitals in $BaFe_2As_2$. The conduction band of primarily Co $3d$ $x^2$-$y^2$ orbital character around the M point is close to a Van Hove singularity, which promotes ferromagnetic low-energy spin fluctuations. The low-energy spin excitations are mainly contributed by the Co $3d$ $x^2$-$y^2$ orbital. The combined effects of strong electronic correlation strength and high density of states at the Fermi level (due to a Van Hove singularity just above the Fermi level) of the Co $3d$ $x^2$-$y^2$ orbital are responsible for the unique A-type AFM order observed in $CaCo_2As_2$. Despite substantial ferromagnetic low-energy spin excitations, $BaCo_2As_2$ and $SrCo_2As_2$ remain paramagnetic down to very low temperature because the Co $3d$ $x^2$-$y^2$ orbital has weaker electronic correlation strength and is further away from the Van Hove singularity. The unique magnetic properties of $CaCo_2As_2$ originate from the increased nearest neighbor Co-Co distance and the significantly reduced As height from the Co plane.

**Acknowledgements**

This work was supported by the Fundamental Research Funds for the Central Universities (Grant No.310421113), the National Natural Science Foundation of China (Grant No. 11674030), the National Key Research and Development Program of China through Contract No. 2016YFA0302300, the National Youth Thousand-Talents Program of China, and the start-up funding of Beijing Normal University. The calculations used high performance computing clusters at Beijing Normal University in Zhuhai and the National Supercomputer Center in Guangzhou.

[*]Email address: yinzhiping@bnu.edu.cn